\documentclass[fleqn,usenatbib,letters]{mnras}

\usepackage{newtxtext,newtxmath}

\usepackage[T1]{fontenc}

\DeclareRobustCommand{\VAN}[3]{#2}
\let\VANthebibliography\thebibliography
\def\thebibliography{\DeclareRobustCommand{\VAN}[3]{##3}\VANthebibliography}

\usepackage{graphicx}	
\usepackage{amsmath}	

\title[WOH\,G64 is still a red supergiant?]{A phoenix rises from the ashes: WOH\,G64 is still a red supergiant, for now\footnote{based on observations made with the Southern African Large Telescope (SALT)}}

\author[J. Th. van Loon \& Keiichi Ohnaka]{
Jacco Th. van Loon,$^{1}$\thanks{E-mail: j.t.van.loon@keele.ac.uk (JvL)}
and Keiichi Ohnaka$^{2}$
\\
$^{1}$Lennard-Jones Laboratories, Keele University, ST5 5BG, UK\\
$^{2}$Instituto de Astrof\'{\i}sica, Departamento de F\'{\i}sica y Astronom\'{\i}a, Facultad de Ciencias Exactas, Universidad Andr\'es Bello, Fern\'andez Concha 700, Las Condes, Santiago, Chile
}

\date{Accepted XXX. Received YYY; in original form ZZZ}

\pubyear{\the\year{}}

\begin{document}
\label{firstpage}
\pagerange{\pageref{firstpage}--\pageref{lastpage}}
\maketitle

\begin{abstract}
For a long time, WOH\,G64 was known as the most extreme red supergiant outside our Galaxy. However, in a matter of years it has faded, its pulsations have become suppressed and the spectrum has become dominated by emission lines from ionised gas, a far cry from the Mira-like pulsation and late M-type spectrum it used to display. Around the same time, a hot dust cloud was discovered using the VLT interferometer. WOH\,G64 has been claimed to have turned into a yellow hypergiant, which could signal a pre-supernova post-red supergiant evolution. Here we present spectra of WOH\,G64 obtained with the Southern African Large Telescope (SALT) between November 2024 and December 2025. Molecular absorption bands from TiO are seen at all times. This implies that WOH\,G64 is currently a red supergiant, and may never have ceased to be. However, the shallow, resolved bands and possible detection of VO hint at a highly extended atmosphere. The continuum appears to be varying, while the line emission shows a different behaviour, suggesting two separate components in the system. Meanwhile, atomic absorption lines are deepening. This places important constraints on scenarios for the dramatic events that are unfolding.
\end{abstract}




\section{Introduction}

Red supergiants are identified with hydrogen-rich core-collapse (type II-P, possibly L) supernova progenitors \citep{Smartt2009} and associated with hydrogen-depleted (type IIb, possibly Ib) core-collapse supernova progenitors following either catastrophic mass loss or envelope stripping by a companion star \citep{Eldridge2013}. SN\,1987A's blue progenitor probably was a red supergiant before \citep{Podsiadlowski1992}. The fate of red supergiants is intimately related to their mass loss \citep[see review by][and references therein]{vanLoon2025}; massive stars are in general affected by binary interaction \citep{Smith2014}.

One of the most extreme red supergiants known, WOH\,G64 was discovered as an optically unremarkable cool giant star in the LMC \citep{Westerlund1981} but subsequently promoted to the very dusty late-M type red supergiant IRAS\,04553$-$6825 \citep{Elias1986}. It exhibited strong Mira-like pulsation and maser emission from a wind \citep{Wood1992,vanLoon1996,vanLoon1998,Whitelock2003,Marshall2004}. Its optical brightness and periodic variability diminished, slowly at first in the 2010s but more dramatically into the 2020s, accompanied by dust formation \citep{Ohnaka2024} and dominance of line emission \citep{MunozSanchez2024}.

We here present spectroscopic observations of WOH\,G64 obtained with the Southern African Large Telescope (SALT) in 2024 and 2025. These optical spectra show emission lines, and a varying continuum characterised by shallow but undisputable molecular absorption that is reminiscent of a red supergiant. We offer an interpretation of these new observations within a scenario of binary star interaction, obviating the need for an imminent supernova explosion.

\section{Observations with SALT}

SALT \citep{Buckley2006} is an effectively 9-m aperture telescope situated at an altitude of 1798\,m on the South African Astronomical Observatory near Sutherland in the Karoo. We used its Robert Stobie Spectrograph \citep[RSS;][]{Burgh2003,Kobulnicky2003} to obtain optical long-slit low-resolution spectra as well as SDSS-r band photometry with the SALTICAM \citep{ODonoghue2006} during acquisition (programmes 2024-2-MLT-005, 2024-2-SCI-020, 2025-2-MLT-007; PI: Jacco van Loon). Some pertinent details are logged in table\,\ref{tab:log}. The position angle of the slit was $28^\circ$ measured east from north, except for observation \#2 when it was set to $30^\circ$.

\begin{table*}
\centering
\caption{Log of long-slit spectroscopy with SALT. Spectral resolution (Full-Width at Half-Maximum) is measured from the 683\,nm telluric O$_2$ emission line.}
\label{tab:log}
\begin{tabular}{lcrcccccccc}
\hline
\# & dd/mm/yy & t (s) & moon phase & airmass & seeing & slit & grating & filter & resolution (nm) & spectral coverage (nm) \\
\hline
1 & 13/11/24 & 900 & 0.88 & 1.237 & $2\rlap{.}^{\prime\prime}3$ & $1\rlap{.}^{\prime\prime}25$ & PG0700 & PC03400 & 0.62 & 359--487 \ \ 495--619 \ \ 627--746 \\
2 & 07/12/24 & 3$\times$900 & 0.43 & 1.265 & $1\rlap{.}^{\prime\prime}0$ & $1\rlap{.}^{\prime\prime}50$ & PG0900 & PC04600 & 0.58 & 590--689 \ \ 695--791 \ \ 796--889 \\
3 & 05/01/25 & 900 & 0.39 & 1.240 & $1\rlap{.}^{\prime\prime}9$ & $1\rlap{.}^{\prime\prime}25$ & PG0700 & PC03400 & 0.67 & 358--486 \ \ 495--619 \ \ 627--746 \\
4 & 02/02/25 & 900 & 0.24 & 1.249 & $1\rlap{.}^{\prime\prime}9$ & $1\rlap{.}^{\prime\prime}25$ & PG0700 & PC03400 & 0.67 & 358--487 \ \ 495--619 \ \ 627--746 \\
5 & 27/10/25 & 900 & 0.34 & 1.265 & $1\rlap{.}^{\prime\prime}4$ & $1\rlap{.}^{\prime\prime}00$ & PG0900 & PC04600 & 0.36 & 645--743 \ \ 750--845 \ \ 850--941 \\
6 & 09/11/25 & 900 & 0.83 & 1.239 & $1\rlap{.}^{\prime\prime}1$ & $1\rlap{.}^{\prime\prime}25$ & PG0700 & PC03400 & 0.59 & 359--487 \ \ 495--619 \ \ 627--747 \\
7 & 27/11/25 & 900 & 0.47 & 1.296 & $3\rlap{.}^{\prime\prime}3$ & $1\rlap{.}^{\prime\prime}00$ & PG0900 & PC04600 & 0.37 & 645--743 \ \ 750--844 \ \ 850--941 \\
8 & 28/11/25 & 900 & 0.58 & 1.260 & $1\rlap{.}^{\prime\prime}5$ & $1\rlap{.}^{\prime\prime}25$ & PG0700 & PC03400 & 0.60 & 358--486 \ \ 494--619 \ \ 626--746 \\
9 & 18/12/25 & 900 & 0.01 & 1.242 & $1\rlap{.}^{\prime\prime}7$ & $1\rlap{.}^{\prime\prime}25$ & PG0700 & PC03400 & 0.62 & 358--486 \ \ 494--619 \ \ 627--746 \\
\hline
\end{tabular}
\end{table*}

Three CCDs cover the spectrum, with small gaps between them, and they were binned by a factor two resulting in a spatial scale of $0\rlap{.}^{\prime\prime}253$\,pix$^{-1}$ (a factor four, $0\rlap{.}^{\prime\prime}507$\,pix$^{-1}$ for the acquisition images) and a dispersion scale of $\approx0.123$\,nm\,pix$^{-1}$ and $\approx0.094$\,nm\,pix$^{-1}$ for the PG0700 and PG0900 gratings, respectively. The detectors were used in normal, faint gain mode (1.0 count per electron) and slow read-out mode (3.3 electrons noise) (fast read-out for acquisition).

The frames were corrected for gain, cross-talk, bias electronic offset; mosaicked, replacing bad pixels; and cleaned of cosmic rays. 2D wavelength calibration was performed based on 2-s exposures of argon arc lamps (neon and xenon for the first and second PG0900 spectra). Spectra were extracted according to the variance-weighted spatial profile using the IRAF routine {\sc apall} after first subtracting the background measured tightly around WOH\,G64 and fitted with a linear function using the IRAF routine {\sc background}. Profile tracing was needed for observations \#2 (of which the three frames were summed), \#6 (which was out of focus in the red), \#7, \#8 and \#9, based on the trace from star\,1 (see below) when WOH\,G64 lacked flux in the blue part. The spectra are badly affected by fringing beyond about $\lambda>800$\,nm, which cancels out after heavy smoothing.

The spectra were divided by a single function that captures the instrumental throughput and response as well as atmospheric and interstellar attenuation, determined from the well exposed spectrum of star\,1. Gaia DR3 \citep{Gaia2023} lists proper motions and parallax consistent with LMC membership, and stellar parameters $T_\mathrm{eff}=4673$\,K, $\log g=2.4$, [Fe/H]$\ =-0.1$ (K-type giant). We used the ATLAS9 model from \citet{Castelli2003} for $T_\mathrm{eff}=4750$\,K, $\log g=2.5$ and solar metallicity, converted it to $F_\lambda$ and divided the spectrum of star\,1 by it. We fitted the result with an $8^\mathrm{th}$-order cubic spline, iterating to reject outliers and avoiding telluric absorption around 680--700 and 758--770\,nm (O$_2$) and 715-730\,nm (H$_2$O), and then normalised it to the r-band (550-690\,nm) level.

\begin{figure}
\hspace{-3mm}\includegraphics[width=90mm]{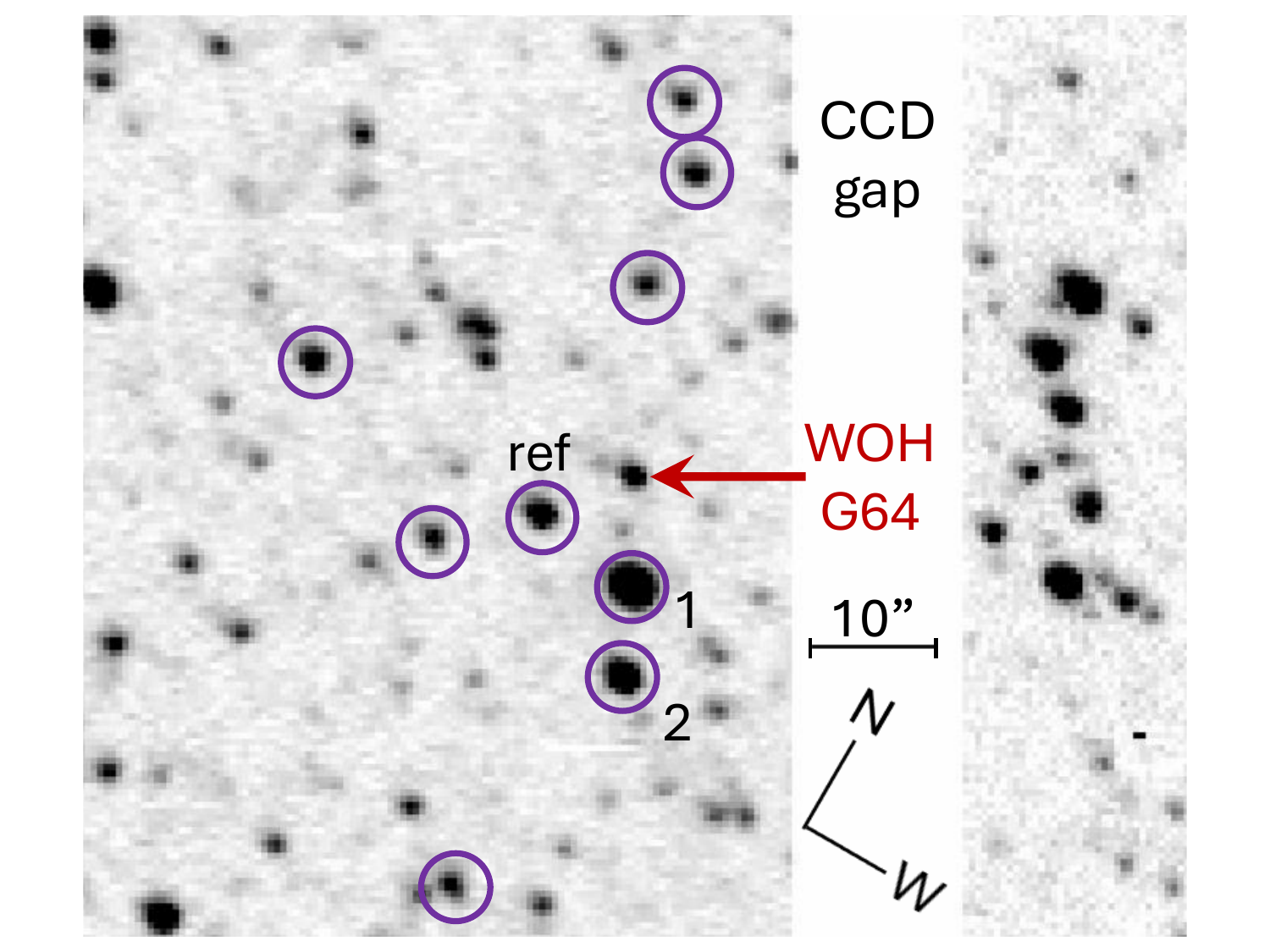}
\caption{SALTICAM acquisition SDSS-r band image from observation \#2. Stars used for photometry are circled; stars 1 \& 2 fall on the spectrograph slit. The photometric primary reference star is labeled with "ref".}
\label{fig:acquisition}
\end{figure}

The acquisition image of observation \#2 is shown in figure\,\ref{fig:acquisition}, where we labelled the reference stars for SDSS-r band photometry and stars 1 and 2 that fall on the spectrograph slit. The r-band magnitudes are taken from the COSMIC-L catalogue \citep{Franco2025} based on Dark Energy Camera observations obtained with the Blanco 4-m telescope at Cerro Tololo on 16 February 2018 (Antonio Franco, private communication). Aperture photometry was performed using the {\sc qphot} routine in IRAF, using a 3-pixel radius aperture and a 2-pixel wide annulus at 5-pixel distance for sky determination. We adopt the relative photometry with respect to the nearest star, labeled "ref" in figure\,\ref{fig:acquisition} as it is of comparable brightness to WOH\,G64. The resulting relative photometry is stable against the average of the others within $\pm0.05$\,mag, but the absolute calibration is uncertain by $\pm0.10$\,mag. WOH\,G64 was $r=15.34$\,mag in February 2018 but some two magnitudes fainter during our observations (table\,\ref{tab:measurements}).

\begin{table*}
\centering
\caption{Measurements of acquisition photometry and spectroscopic features. The H$\alpha$+[N\,{\sc ii}] contrast is with respect to the local continuum (see section~3.3).}
\label{tab:measurements}
\begin{tabular}{lcccccccccc}
\hline
\# & JD & SDSS-r & TiO\,706\,nm & H$\alpha$+[N\,{\sc ii}] & [Ca\,{\sc ii}] & [S\,{\sc ii}] / [O\,{\sc i}] & [O\,{\sc i}] / [N\,{\sc i}]  & K\,{\sc i}\,766+770 & O\,{\sc i}\,777 & Na\,{\sc i}\,819 \\
 & (day) & (mag) & (band head) & (line contrast) & EW (\AA) & (flux ratio) & (flux ratio) & EW (\AA) & EW (\AA) & EW (\AA) \\
\hline
1 & 2460628 & $17.86\pm0.05$ & $1.20\pm0.11$ & $2.11\pm0.19$ & $-4.6\pm0.6$ & $1.51\pm0.18$ & $0.94\pm0.08$ & & & \\
2 & 2460652 & $17.82\pm0.05$ & $1.19\pm0.04$ & $2.21\pm0.07$ & $-6.3\pm0.5$ & & & $1.29\pm0.29$ & $0.73\pm0.11$ & $2.0\pm0.2$ \\
3 & 2460681 & $17.72\pm0.05$ & $1.20\pm0.05$ & $1.86\pm0.12$ & $-5.9\pm1.0$ & $1.74\pm0.13$ & $1.05\pm0.08$ & & & \\
4 & 2460709 & $17.62\pm0.05$ & $1.21\pm0.06$ & $1.39\pm0.08$ & $-4.2\pm0.7$ & $1.60\pm0.10$ & $0.91\pm0.04$ & & & \\
5 & 2460976 & $17.27\pm0.05$ & $1.17\pm0.04$ & $1.08\pm0.08$ & $-3.0\pm0.7$ & $2.85\pm0.20$ & & $2.06\pm0.14$ & $1.63\pm0.42$ & $2.6\pm0.5$ \\
6 & 2460989 & $17.35\pm0.05$ & $1.15\pm0.05$ & $1.19\pm0.07$ & $-3.6\pm1.0$ & $1.75\pm0.12$ & $1.11\pm0.08$ & & & \\
7 & 2461007 & $17.36\pm0.05$ & $1.18\pm0.07$ & $0.98\pm0.08$ & $-2.3\pm0.4$ & & & $2.10\pm0.14$ & $1.24\pm0.18$ & $3.6\pm0.2$ \\
8 & 2461008 & $17.39\pm0.05$ & $1.19\pm0.03$ & $1.16\pm0.08$ & $-2.4\pm0.5$ & $1.66\pm0.11$ & $1.03\pm0.08$ & & & \\
9 & 2461028 & $17.38\pm0.05$ & $1.23\pm0.06$ & $1.19\pm0.06$ & $-2.3\pm0.4$ & & & \\
\hline
\end{tabular}
\end{table*}

\section{Spectral analysis}

\begin{figure*}
\hspace{-1mm}\includegraphics[width=179mm]{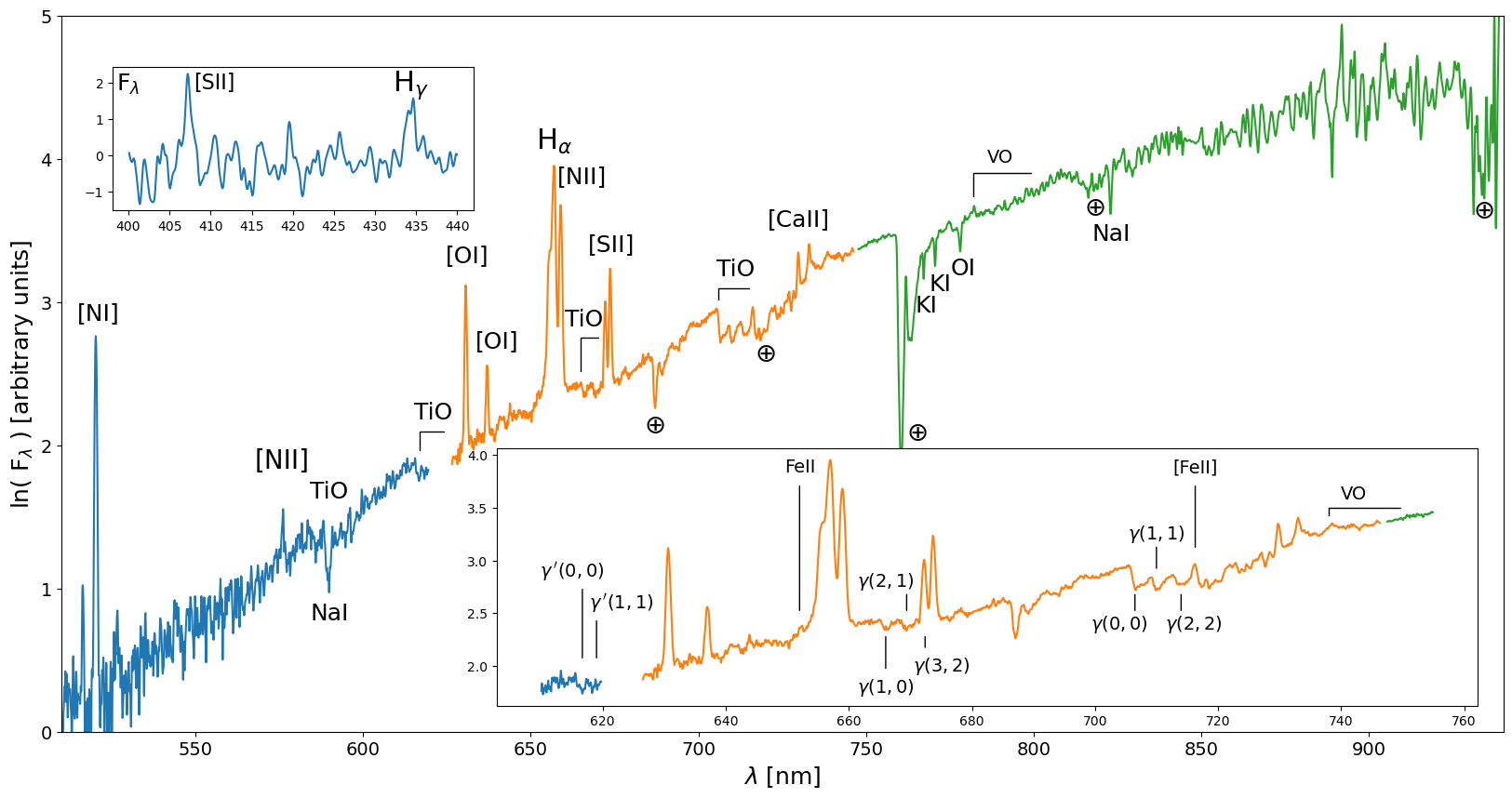}
\caption{Grand total of nine SALT spectra of WOH\,G64, representative of the spectral appearance during the period November 2024 -- December 2025. Spectra below 620\,nm and beyond 748\,nm have been smoothed with a two-pixel Gaussian kernel except in the large inset (where TiO transitions are identified).}
\label{fig:spectrum}
\end{figure*}

To aid in obtaining an overview of the optical spectrum and setting the scene for the more detailed analysis described below, figure\,\ref{fig:spectrum} shows a holistic combination of all SALT spectra. The individual spectra were resampled onto a common dispersion axis and scaled to a common median flux between 610--710\,nm so they could be stitched together. It should be understood that this does not represent an accurate reflection of the full spectrum at any one time, but it is still meaningful since we do not observe dramatic changes.

\subsection{Atomic and ionic line emission}

The optical spectrum of WOH\,G64 at first glance is dominated by several strong emission lines on top of a red pseudo-continuum. The strongest is H$\alpha$, flanked by [N\,{\sc ii}]; H$\beta$ fell on a CCD gap but H$\gamma$ is detected (there is virtually no continuum emission below $\lambda<500$\,nm). Case B recombination yields $I(\mathrm{H}\alpha)/I(\mathrm{H}\gamma)=6.08$; the observed ratio $\sim20$ implies a differential reddening of a factor 3.3 between 0.43\,$\mu$m and 0.66\,$\mu$m -- or $E(B-R)\sim1.3\equiv A_V\sim2.6$\,mag.

The [S\,{\sc ii}] doublet ratio at 672/673\,nm implies an electron density $n_\mathrm{e}\sim10^3$\,cm$^{-3}$ \citep{Osterbrock2006}; the line ratio is reversed compared to that in the adjacent tenuous ISM. Compared to typical H\,{\sc ii} regions, relatively high densities are also implied by the strong [N\,{\sc i}] 520\,nm and [O\,{\sc i}] 630/636\,nm lines. We note that all the above lines were already present in $20^\mathrm{th}$-century spectra of WOH\,G64.

The 2016 spectrum presented by \citet{MunozSanchez2024} showed additional low-excitation lines such as Fe\,{\sc i}, Fe\,{\sc ii}, [Fe\,{\sc ii}] and [Ca\,{\sc ii}] 729/732\,nm as well as the 850/854/866-nm calcium triplet in emission. These seemed to have diminished in their 2021 spectrum; in our low-resolution spectra the calcium triplet is not seen (its emission and absorption may cancel out in our low-resolution spectra) but the [Ca\,{\sc ii}] doublet and [Fe\,{\sc ii}] 716\,nm persist, with a hint of Fe\,{\sc ii} 652\,nm.

\subsection{Molecular band absorption}

The strongest molecular absorption in the optical spectrum of an M-type star is due to electronic transitions from the ground state of titanium oxide (TiO), X\,$^3\Delta\rightarrow$\,C\,$^3\Delta$ ($\lambda(\alpha)=517$\,nm), $\rightarrow$\,B\,$^3\Pi$ ($\lambda(\gamma^\prime)=616$\,nm) and $\rightarrow$\,A\,$^3\Phi$ ($\lambda(\gamma)=706$\,nm), where the wavelengths are the heads of red-asymmetric bands \citep{Kaminski2009}. Additional bands between low-lying vibrational levels within these electronic transitions are seen, for instance $\gamma^\prime(1,0)$ around 585\,nm and $\gamma(1,0)$ around 665\,nm -- these bands are often lacking sharp edges because of the overlap from other vibrational sub-structure.

All spectral settings in our SALT observations of WOH\,G64 cover the $\gamma(0,0)$ band with an edge at 706\,nm. It is seen in all spectra; the bands at 585, 616 and 665\,nm are also seen. These are not confused by telluric absorption from O$_2$ (band heads at 687 and 760\,nm) and H$_2$O (rounder bands around 720, 820 and 935\,nm). While the blue component of [S\,{\sc ii}] sits on top of the $\gamma(3,2)$ transition this cannot completely account for its positive red/blue line ratio.

We see hints of the B\,$^4\Pi\rightarrow$\,X\,$^4\Sigma^-$ (0,0) and (1,0) band heads of vanadium oxide (VO) around 782\,nm and 738\,nm, respectively; these are among the strongest VO bands and not confused by any TiO \citep{Kaminski2009}. This would suggest a late-M type spectrum. However, the TiO bands are much shallower than normally observed in cool star photospheres and they are resolved into their ro-vibrational sub-structure. This is more typical of the rarefied, cold pseudo-photospheres overlying some post-asymptotic giant branch stars such as U\,Equ and IRAS\,08182--6000 \citep{Couch2003} and the post-merger object V838\,Mon \citep{Liimets2023}. In hindsight, pre-2010 spectra \citep{Levesque2009} show hints of this, pointing at a high $L/M$ ratio following the sustained, heavy mass loss over its life as a red supergiant \citep{vanLoon2005,Ohnaka2008}.

\subsection{Searching for signs of spectral evolution and complexity}

To investigate possible temporal changes in the TiO absorption we measured the strength of the 706\,nm band head from the ratio of the mean flux in 1-nm regions around 705.0 and 707.5\,nm. For reference, the value in the grand total spectrum (figure\,\ref{fig:spectrum}) is $1.190\pm0.016$, a highly significant detection. There is marginal evidence for the band to have weakened between early-2025 and late-2025 (table\,\ref{tab:measurements}) but to have since strengthened again.

\begin{figure}
\hspace{0mm}\includegraphics[width=85mm]{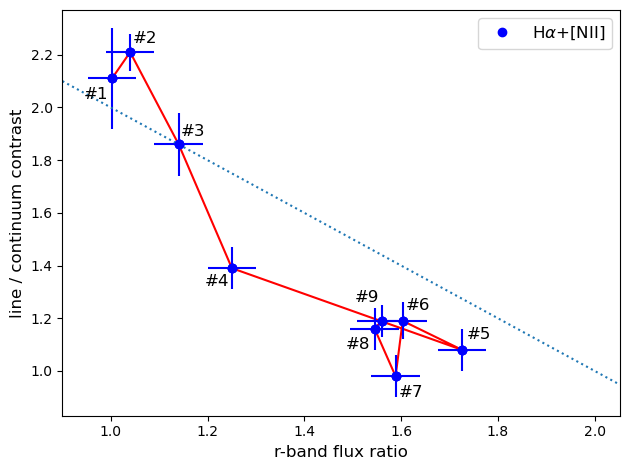}
\caption{Contrast between H$\alpha$ + [N\,{\sc ii}] emission line complex and continuum (section 3.3) versus relative flux in the r-band. The dotted line is for the case that the line complex remained absolutely as bright as in observation \#3.}
\label{fig:linecontinuum}
\end{figure}

We also quantify the contrast of the H$\alpha$+[N\,{\sc ii}] emission complex, relative to the continuum it sits on, by dividing its continuum-subtracted mean level over 653--660\,nm by the mean continuum level interpolated in this same range. This means that the mean emission in the complex is as bright as the underlying continuum, when this ratio is unity. There is a striking decline in time (table\,\ref{tab:measurements}), which strongly correlates with the brightening in the SDSS-r band (550--690\,nm). In figure\,\ref{fig:linecontinuum} the r-band magnitudes are converted to in-band flux relative to the first observation. The trend can be explained by the continuum brightening but the line emission remaining more stable.

Instead, absorption lines from neutral metals have markedly strengthened over the year of observations, by about a factor 1.6--2 for the K\,{\sc i}\,766+770 doublet, the O\,{\sc i}\,777 triplet and the Na\,{\sc i}\,819 doublet -- which had EW $=1.51\pm0.04$\,\AA\ in 1995/1996 \citep{vanLoon2001}. The 589\,nm doublet of Na\,{\sc i} was difficult to measure but also showed signs of increasing from $2.9\pm0.5$\,\AA\ in observation \#3 to $5.7\pm1.1$\,\AA\ in observations \#6 and \#8. It remains to be seen whether this trajectory towards a yellow hypergiant spectrum continues.

\section{An emerging scenario of binary interaction}

The first conclusion that can be drawn from our analysis is that there are two components in this system. Component A is  identified with the continuum of the red supergiant, whilst the source of the line emission resides in component B -- which must be UV-bright in order to ionise hydrogen, and irradiate relatively dense gas: [N\,{\sc i}] 520\,nm, [O\,{\sc i}] 630/636\,nm and [Ca\,{\sc ii}] 729/732\,nm are commonly found in the inner disk of B[e] stars \citep{Aret2016}. They have been detected in the LMC before \citep[LI-LMC\,1522:][]{vanLoon2005} and in the post-red supergiant binary AFGL\,4106 \citep{vanLoon1999a}. Like WOH\,G64, LI-LMC\,1522 shows a composite spectrum of line emission and a cool continuum \citep{vanLoon2005}.

The decline in optical brightness in the 2010s and the discovery of a fresh dust cloud that could have started to form around that time \citep{Ohnaka2024}, along with the spectral dominance by emission lines suggest that the red supergiant, WOH\,G64\,A had been obscured more than the hot component, WOH\,G64\,B. The Balmer decrement does imply that WOH\,G64\,B is also affected by dust, either from the pre-2008 envelope of WOH\,G64\,A at $>15R_\mathrm{A}$ \citep{Ohnaka2008} or from the newly formed dust at $\sim10R_\mathrm{A}$ \citep{Ohnaka2024}.

\begin{figure}
\hspace{0mm}\includegraphics[width=85mm]{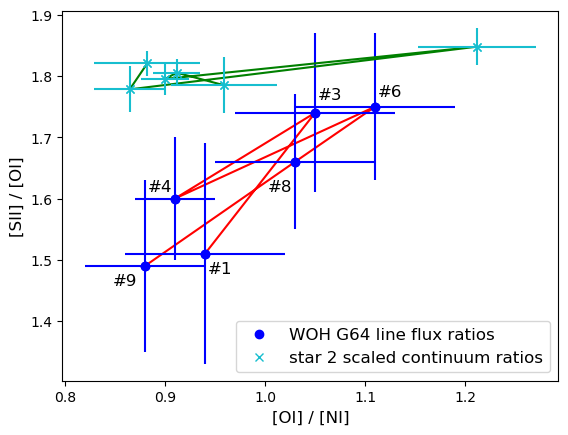}
\caption{Emission line flux ratios of [S\,{\sc ii}]\,672+673\,nm / [O\,{\sc i}]\,630\,nm versus [O\,{\sc i}]\,630\,nm / [N\,{\sc i}]\,520\,nm. While they change in concert, they go back and forth between epochs, consistent with time-varying reddening by dust. Apart from one outlier, the equivalent continuum ratios (scaled $\div1.7$ along $x$ and $\times1.7$ along $y$ to aid comparison) for star 2 do not exhibit the same behaviour.}
\label{fig:lineratios}
\end{figure}

We detect a tentative correlation between the flux ratios of the [S\,{\sc ii}]\,672+673\,nm and [O\,{\sc i}]\,630\,nm on the one hand, and [O\,{\sc i}]\,630\,nm and [N\,{\sc i}]\,520\,nm on the other, in a way that is suggestive of variations in reddening by dust (Fig.~\ref{fig:lineratios}). This may also explain the faster decline in H$\alpha$+[N\,{\sc ii}] emission than the continuum increased (Fig.~\ref{fig:linecontinuum}), and a similar decline in the equivalent width (EW) of the [Ca\,{\sc ii}]\,729+732\,nm emission (table\,\ref{tab:measurements}).

To test the flux ratios are not due to chromatic slit losses we also measured the continuum flux ratios of the equivalent spectral regions 670--675\,nm (pseudo-[S\,{\sc ii}]), 628--633\,nm (pseudo-[O\,{\sc i}]) and 518--523\,nm (pseudo-[N\,{\sc i}]) for star 2 which should be affected more since it was not centred in the slit. With the exception of observation \#6 (with the best seeing, exacerbating any chromatic losses if the star was placed near the edge of the slit) the variations are much smaller than those seen in WOH\,G64.

The latter are not monotonic in time, suggesting the expanding dust cloud may be blobby on au scales (timescales of months at a wind speed $\sim25$\,km\,s$^{-1}$ -- \citet{vanLoon1996,Marshall2004}). This would mesh well with WOH\,G64\,A's cool atmosphere being patchy, resulting in the weak TiO bands and exposing a warmer photosphere exemplified by strengthening atomic absorption. If it is anything to go by, U\,Equ went on to metamorphose into the hot end product of intermediate-mass stellar evolution \citep{Kaminski2024}. On the other hand, the continuum and molecular absorption appear to vary as expected for a slowly (multi-year) pulsating atmosphere. WOH\,G64 may yet be clinging on to its red supergiant modality.

We need to explain the shortening pulsation period, though, from $>900$\,d in the 1980s \citep{Wood1992} to $<900$\,d around the millennium \citep{Whitelock2003}, and the slow warming from spectral type M7.5 in the 1990s to M5 a decade hence \citep{Levesque2009}. It implies a shrinking of the photosphere of WOH\,G64\,A, and is difficult to reconcile with the formation of fresh dust. The timeline of events also suggests the shrinking preceded the obscuration.

We here propose that an approach of WOH\,G64\,B could have led to the tightening of the Roche Lobe around WOH\,G64\,A. It would have been a slow process, since a radius $R_\mathrm{A}\sim1540$\,R$_\odot$ \citep{Levesque2009} and assuming a current mass $M_\mathrm{A}\sim20$\,M$_\odot$ imply orbital speeds $\ll50$\,km\,s$^{-1}$ taking seven decades to traverse a semi-circle at $10R_\mathrm{A}$. The shallower gravitational potential due to the presence of WOH\,G64\,B would stretch the already bloated atmosphere of WOH\,G64\,A. This would increase the circumstellar gas density, triggering the formation of the elongated dust cloud \citep{Ohnaka2024}. Some of this gas might feed the disk around WOH\,G64\,B, which however has shown little sign of being affected by the interaction -- save for a marginal increase in [S\,{\sc ii}]\,673/672 line ratio (and thus of density) from $1.55\pm0.09$ in the first four observations to $1.67\pm0.05$ in the last five observations. Meanwhile, the increasingly transparent expanding atmosphere of WOH\,G64\,A causes the layer of optical depth $\tau\sim1$ to retreat to a deeper, warmer photosphere, and the star exhibiting faster but shallower radial pulsation.

It is likely the orbital plane of the binary is at a small angle to the plane of the sky, because there are no credible orbital motions recorded \citep[offsets can be explained by pulsation or scattering off receding dust --][]{vanLoon1998}, no reported eclipses or transits, and it explains the morphology of the dust envelope as viewed from Earth \citep{Roche1993,vanLoon1999b,Ohnaka2008}. The orbital period must exceed a century, as no similar events are known to have occurred since the 1950s.

The alternative, a Betelgeuse-like Great Dimming \citep{Dupree2025} due to convection-driven ejection and cooling-induced dust formation, seems less likely. Such event tends to last of order the dynamical timescale of pulsation and convection, but in the case of WOH\,G64 the changes have taken place over several pulsation periods. A merger scenario can also be ruled out as no vast amount of gravitational energy has been liberated -- the bolometric luminosity $\sim3\times10^5$\,L$_\odot$ has stayed stable suggesting the persistence of the supergiant \citep{Ohnaka2024,MunozSanchez2024}.

\section{Conclusions}

We have presented evidence that the remarkable changes witnessed in the $21^\mathrm{st}$-century in the optical brightness and spectrum of the most extreme known extragalactic red supergiant, WOH\,G64 may be due to binary interaction. While the line emission of hot gas associated with component B outshone the obscured light from component A, the latter has kept its M-type appearance and therefore its red supergiant status. However, the cool atmosphere appears highly extended and may soon reveal a hotter descendant. We may be witnessing the birth of a hydrogen-poor core-collapse supernova progenitor, unless it re-establishes itself as the pulsating red supergiant as it was known.

\section*{Acknowledgements}

We thank the referee, Tomek Kami\'nski for his insightful report. All of the observations reported in this paper were obtained with the Southern African Large Telescope (SALT). IRAF is distributed by the National Optical Astronomy Observatory, which is operated by the Association of Universities for Research in Astronomy (AURA) under a cooperative agreement with the National Science Foundation. This research has made use of the VizieR catalogue access tool, CDS, Strasbourg, France
\citep{vizier2000}. KO acknowledges the support of the Agencia Nacional de Investigaci\'on y Desarrollo (ANID) through the FONDECYT Regular grant 1240301.

\section*{Data Availability}

The spectroscopic data including the acquisition images become available to the public on the SALT archive at https://ssda.saao.ac.za/ after the proprietary period has expired.



\bibliographystyle{mnras}
\bibliography{WOHG64} 








\bsp	
\label{lastpage}
\end{document}